\documentclass[acmsmall,screen]{acmart}

\usepackage{packages}

\begin{document}

\settopmatter{printacmref=false}
\setcopyright{none}
\renewcommand\footnotetextcopyrightpermission[1]{}

\renewcommand{\authorsaddresses}{}

\renewcommand{\acmConference}[3]{}
\renewcommand{\acmBooktitle}[1]{}
\renewcommand{\acmYear}[1]{}
\renewcommand{\acmDOI}[1]{}
\renewcommand{\acmISBN}[1]{}

\title{TriCalRAG: A Three-Strategy, Retrieval-Augmented Bench for On-Prem LLM-Based Root Cause Analysis in AIOps}

\author{Rohit Patel}
\correspondingauthor
\authornotemark[1]
\email{g25ait2089@iitj.ac.in}
\orcid{1234-5678-9012}
\affiliation{%
  \institution{Indian Institute of Technology Jodhpur}
  \city{Jodhpur}
  \state{Rajasthan}
  \country{India}
}

\author{Susil Kumar Mohanty}
\email{susilmohanty@iitj.ac.in}
\orcid{1234-5678-9012}
\affiliation{%
  \institution{Indian Institute of Technology Jodhpur}
  \city{Jodhpur}
  \state{Rajasthan}
  \country{India}
  }

\author{Jeenal Chaudhary}
\email{g25ait2027@iitj.ac.in}
\orcid{1234-5678-9012}
\affiliation{%
  \institution{Indian Institute of Technology Jodhpur}
  \city{Jodhpur}
  \state{Rajasthan}
  \country{India}
  }

\renewcommand{\shortauthors}{Patel et al.}

\begin{abstract}
AIOps platforms generate massive volumes of heterogeneous logs, making timely and reliable root cause analysis (RCA) increasingly challenging. While cloud-hosted large language models (LLMs) offer promising capabilities for automated incident diagnosis, their dependence on external services raises concerns about privacy, compliance, latency, and inference cost, limiting their practicality for continuous, large-scale deployment. Moreover, aggregate detection metrics can conceal prediction bias, making accuracy alone insufficient to assess reliability. To address these challenges, we present TriCalRAG, a reproducible benchmark for evaluating privacy-aware, resource-efficient on-premise LLM-based log-anomaly detection with generated root-cause explanations across four real-world datasets: BGL, HDFS, Thunderbird, and OpenStack.

The primary evaluation compares Qwen2.5-14B and Mistral-Small-22B, served through vLLM on one NVIDIA RTX PRO 6000 GPU (96 GB), under zero-shot, few-shot, and retrieval-augmented generation (RAG) prompting across three
data-sampling seeds. We report F1, bootstrap confidence intervals, predicted-positive rates, throughput, and memory use, with DeepLog as a held-out classical baseline. Mistral-Small attains a higher macro-averaged F1 than Qwen2.5-14B (0.644 versus 0.560), whereas Qwen provides approximately twice the throughput. A separate log-probability evaluation compares raw decisions with Contextual Calibration (CC) and Batch Calibration (BC): neither correction consistently improves prediction-balance diagnostics across prompting strategies. Supplementary single-run comparisons extend evaluation to 4-bit Llama-3.1-70B via local Ollama and Claude Haiku 4.5 via Anthropic's cloud API; RAG improves F1 on all four datasets for both models. Claude's reported aggregate F1 increases from 0.566 to 0.695, while the estimated API cost rises from \$0.94 to \$2.15 per 1,000 incidents. Local deployment ablations show approximately 41-fold throughput scaling with batching and 20\% lower latency with 4-bit quantization on the tested workload. These findings support retrieval as a useful context for anomaly decisions, while the supplementary protocols, unvalidated explanation quality, and prediction-balance diagnostics limit broader claims about RCA accuracy and probabilistic calibration.

These results indicate that evaluating LLM-based AIOps systems requires reporting prediction-balance and deployment-efficiency diagnostics alongside conventional detection metrics. We release the benchmark harness, evaluation code, and reproducibility artifacts to support further research on reliable on-premise LLM-based AIOps.
\vspace{0.5 cm}
\end{abstract}



\maketitle

\vspace{0.5 cm}

\section{Introduction}
\label{sec:intro}

Modern IT infrastructures continuously generate large volumes of
heterogeneous operational logs, making timely incident diagnosis and
root-cause analysis (RCA) increasingly difficult for on-call engineers.
This challenge has motivated the adoption of AIOps~\cite{dang2019aiops}
and, more recently, large language models (LLMs) for automated incident
analysis. Unlike conventional anomaly detectors~\cite{chandola2009anomaly},
which typically reduce an incident to a binary anomalous/normal decision,
LLM-based approaches can additionally generate natural-language
root-cause explanations and remediation suggestions.

However, evaluating LLM-based RCA introduces two practical challenges.
First, many LLM-based systems rely on cloud-hosted APIs. Production logs
may contain sensitive infrastructure details, internal hostnames,
credentials, or customer-related information, making external processing
undesirable in privacy- or compliance-sensitive environments. Cloud API
use also introduces per-query inference cost and network latency that can
grow with operational workload.

Second, conventional aggregate detection metrics can obscure important
prediction behavior. In a balanced anomaly-detection benchmark, a model
that predicts ``anomaly'' for nearly every incident can achieve high
recall and a moderate F1 score while providing limited discriminative
value. Evaluating only aggregate detection performance can therefore give
an incomplete picture of operational reliability.

These observations motivate two central questions:
\begin{enumerate}[label=\textbf{Q\arabic*:}]
    \item \textbf{Can useful LLM-based log analysis be performed
locally on a single machine GPU?}

    \item \textbf{How should LLM-based RCA systems be evaluated when
aggregate detection metrics may hide extreme prediction imbalance?}
\end{enumerate}

We present \textbf{TriCalRAG}, a reproducible benchmark for evaluating
open-weight LLMs served locally via vLLM~\cite{vllm2023} on a single
NVIDIA RTX PRO 6000 (96\,GB VRAM) workstation. TriCalRAG evaluates four
public LogHub datasets~\cite{loghub} under three prompting strategies:
zero-shot, few-shot, and retrieval-augmented generation (RAG) over
labeled historical incidents. The benchmark additionally includes a
DeepLog~\cite{deeplog2017} classical baseline, supplementary evaluations
with Llama-3.1-70B and Claude Haiku 4.5, and deployment-oriented
measurements covering throughput, memory use, quantization, and API cost.
The overall architecture is shown in Fig.~\ref{fig:tricalrag}.

Our contributions are as follows:

\begin{enumerate}

  \item A reproducible benchmark for on-premise LLM-based RCA spanning
        four real log datasets, two primary open-weight models, three
        prompting strategies, and a classical non-LLM baseline, evaluated
        on a single workstation GPU.

  \item A repeated evaluation protocol using three data-sampling seeds
        per primary configuration with bootstrap 95\% confidence
        intervals.

  \item An evaluation showing that F1 alone can obscure extreme
        prediction imbalance, together with predicted-positive rate as
        a complementary diagnostic and an analysis of how RAG affects
        this behavior.

  \item A separate mechanism and generalization study comparing RAG with
        Contextual Calibration and Batch Calibration and extending the
        zero-shot/RAG comparison to Llama-3.1-70B and Claude Haiku 4.5.

  \item A reproducible evaluation harness and deployment study covering
        batching, quantization, throughput, memory use, and cloud API
        cost.

\end{enumerate}

The remainder of this paper is organized as follows.
Section~\ref{sec:related} reviews related work on log anomaly detection,
LLM-based RCA, retrieval-augmented log analysis, LLM serving, and
in-context calibration. Section~\ref{sec:benchmark} presents the
TriCalRAG benchmark design, including datasets, incident construction,
prompting strategies, retrieval, and evaluated systems.
Section~\ref{sec:expsetup} describes the research questions,
experimental protocol, metrics, and deployment setup.
Section~\ref{sec:results} reports the results organized around the four
research questions. Section~\ref{sec:discussion} discusses the practical
implications of the findings. Section~\ref{sec:limitations} presents the
study limitations, and Section~\ref{sec:conclusion} concludes the paper
and outlines future work.

\section{Related Work}
\label{sec:related}

This section reviews work related to automated log analysis and root
cause analysis (RCA), spanning classical anomaly detection, LLM-based
log analysis, retrieval-augmented methods, efficient LLM serving, general
LLM benchmarking, and calibration methods for in-context classification.
We conclude by identifying the gap addressed by TriCalRAG: a unified
evaluation of detection effectiveness, prediction balance, and deployment
efficiency for on-premise LLM-based incident analysis.

\subsection{Log Anomaly Detection}
\label{sec:related-logad}

Classical log anomaly detection learns patterns of normal system behavior
and identifies deviations from those patterns. DeepLog~\cite{deeplog2017}
treats log keys as a sequence and trains an LSTM to predict the next log
key, flagging sequences whose observed continuation falls outside the
model's top-$k$ predictions. LogAnomaly~\cite{loganomaly2019} extends this
idea with template embeddings that capture semantic similarity between
log messages.

These approaches provide effective anomaly detection but generally do
not generate natural-language root-cause explanations or remediation
recommendations, which is one capability explored by recent LLM-based
systems. LogHub~\cite{loghub} provides a widely used collection of
real-world system-log datasets, including the four datasets used in
TriCalRAG. He et al.~\cite{loglizer} provide complementary tools and
benchmarks for automated log parsing, which commonly precedes anomaly
detection. At the broader systems level, Soldani and
Brogi~\cite{soldani2022rca} survey anomaly detection and RCA for
microservice and cloud applications, providing the operational context
in which our benchmark is situated.

\begin{table}[!htbp]
\centering
\caption{Comparison of representative approaches and the scope of
TriCalRAG. A checkmark indicates that the aspect is explicitly addressed;
a dash indicates that it is not a primary focus.}
\label{tab:related-work}

\scriptsize
\setlength{\tabcolsep}{2.5pt}
\renewcommand{\arraystretch}{1.15}

\begin{tabularx}{\columnwidth}{
    >{\raggedright\arraybackslash}p{1.35cm}
    >{\raggedright\arraybackslash}p{2.3cm}
    >{\centering\arraybackslash}p{0.75cm}
    >{\centering\arraybackslash}p{0.58cm}
    >{\centering\arraybackslash}p{0.55cm}
    >{\centering\arraybackslash}p{1.20cm}
    >{\centering\arraybackslash}p{0.55cm}
    >{\centering\arraybackslash}p{0.55cm}
    >{\raggedright\arraybackslash}X
}

\toprule

\textbf{Category}
&
\textbf{Representative work}
&
\textbf{AD}
&
\textbf{RCA}
&
\textbf{RAG}
&
\textbf{On-prem.}
&
\textbf{Calib.}
&
\textbf{Eff.}
&
\textbf{Primary focus}
\\

\midrule

Classical log
detection
&
DeepLog~\cite{deeplog2017};
LogAnomaly~\cite{loganomaly2019}
&
\checkmark
&
$\pm$
&
--
&
\checkmark
&
--
&
\checkmark
&
Sequence-based anomaly detection
\\[3pt]

LLM log
analysis
&
LogGPT~\cite{loggpt2023};
LogLLM~\cite{logllm2024};
LogLM~\cite{liu2024loglm}
&
\checkmark
&
$\pm$
&
--
&
$\pm$
&
--
&
$\pm$
&
LLM-based log analysis
\\[3pt]

LLM-based
RCA
&
Ahmed et al.~\cite{ahmed2023recommending};
GPT-4 RCA~\cite{gpt4rca2024}
&
\checkmark
&
\checkmark
&
--
&
$\pm$
&
--
&
$\pm$
&
Automated incident diagnosis
\\[3pt]

RAG-based
analysis
&
LogRAG~\cite{lograg2024};
EagerLog~\cite{duan2025eagerlog}
&
\checkmark
&
$\pm$
&
\checkmark
&
$\pm$
&
--
&
$\pm$
&
Retrieved contextual evidence
\\[3pt]

LLM
serving
&
vLLM~\cite{vllm2023};
FlashAttention~\cite{dao2022flashattention}
&
--
&
--
&
--
&
\checkmark
&
--
&
\checkmark
&
Efficient local inference
\\[3pt]

LLM
benchmarking
&
MMLU~\cite{hendrycks2021mmlu};
HELM~\cite{liang2022helm};
SWE-bench~\cite{jimenez2023swebench}
&
$\pm$
&
$\pm$
&
$\pm$
&
$\pm$
&
\checkmark
&
\checkmark
&
Multi-axis model evaluation
\\[3pt]

In-context
calibration
&
Contextual Calibration~\cite{zhao2021calibrate};
Batch Calibration~\cite{zhou2024batch}
&
$\pm$
&
--
&
--
&
--
&
\checkmark
&
--
&
Label-bias correction
\\[3pt]

\rowcolor{gray!12}

\textbf{TriCalRAG}
&
\textbf{This work}
&
\checkmark
&
\checkmark
&
\checkmark
&
\checkmark
&
\checkmark
&
\checkmark
&
\textbf{Unified on-premise AIOps evaluation}
\\

\bottomrule

\end{tabularx}

\vspace{1mm}

\noindent
{\footnotesize
AD: Anomaly Detection; RCA: Root Cause Analysis;
Calib.: explicit calibration or label-bias analysis;
Eff.: Inference Efficiency.
$\pm$ denotes Partial or Indirect Coverage.
}

\end{table}

\subsection{LLMs for Log Analysis and RCA}
\label{sec:related-llm}

Recent work applies LLMs across multiple stages of the log-analysis
pipeline; Akhtar et al.~\cite{llmeventlogsurvey2025} survey this area
broadly. LogGPT~\cite{loggpt2023} explores prompting a commercial LLM
for log anomaly detection with structured outputs, similar in format to
our task definition. LogLLM~\cite{logllm2024} and
LogLM~\cite{liu2024loglm} investigate fine-tuned and instruction-tuned
open-weight models for log analysis, while
LLMeLog~\cite{he2024llmelog} enriches log events with LLM-generated
semantic information prior to detection.

ClsLog~\cite{xiao2025clslog} combines large and small models to balance
detection performance against inference cost, while
Anomaly-Gen~\cite{li2025anomalygen} uses an LLM to synthesize training
sequences rather than perform classification directly. At the upstream
log-parsing stage, UniLog~\cite{xu2024unilog},
LogParser-LLM~\cite{zhong2024logparserllm}, and
Lilac~\cite{jiang2024lilac} apply LLMs and in-context learning to log
template extraction. Song et al.~\cite{song2025confront} additionally
study LLM fine-tuning for anomaly detection in behavior logs.

For RCA specifically, Ahmed et
al.~\cite{ahmed2023recommending} conduct a large-scale study of
LLM-generated root-cause and mitigation recommendations across more than
40{,}000 production incidents. Zhang et al.~\cite{gpt4rca2024} apply
in-context learning with GPT-4 to automated root causing of cloud
incidents. These studies demonstrate the potential of LLMs for
operational diagnosis. TriCalRAG addresses a complementary question:
whether such analysis is practical on a single on-premise GPU and how
detection effectiveness should be interpreted when aggregate metrics may
obscure extreme prediction imbalance.

\subsection{Retrieval-Augmented Log Analysis}
\label{sec:related-rag}

Retrieval-augmented generation (RAG) was introduced by Lewis et
al.~\cite{lewis2020rag} as a framework for grounding language-model
generation in retrieved non-parametric context. In log analysis,
LogRAG~\cite{lograg2024} retrieves semantically similar historical log
templates as context for LLM-based anomaly detection in a semi-supervised
setting. EagerLog~\cite{duan2025eagerlog} combines active learning with
retrieval augmentation to reduce labeling requirements.

TriCalRAG differs from these approaches in three main respects.
First, we retrieve complete labeled incidents, including their
ground-truth outcome, rather than individual log templates. Second, RAG
is evaluated as one of three controlled prompting strategies rather than
as a standalone method. Third, we examine not only detection
effectiveness but also how retrieval affects prediction balance and
deployment behavior.

Prior work on RAG pipeline robustness,
TriShieldRAG~\cite{rohitpatel2026trishieldrag}, studies
knowledge-corruption risks in retrieval-augmented systems. Although that
work addresses adversarial robustness rather than RCA, its analysis of
retrieval behavior provides relevant context for reasoning about
retrieved incident precedent in operational settings.

\subsection{LLM Serving Systems}
\label{sec:related-serving}

TriCalRAG uses vLLM~\cite{vllm2023}, whose PagedAttention memory
management supports high-throughput batched inference and makes
single-device LLM serving practical. Related efficiency techniques
include IO-aware attention implementations such as
FlashAttention~\cite{dao2022flashattention} and parameter-efficient
fine-tuning methods such as LoRA~\cite{hu2022lora}. Although these
techniques are not all evaluated directly in our benchmark, they reflect
the broader effort to reduce the computational cost of deploying large
language models. This issue becomes increasingly important as
open-weight model sizes continue to grow~\cite{kaplan2020scaling}.

Our focus differs from general LLM-serving research: rather than
optimizing the serving engine itself, we measure the detection and
deployment behavior achievable using a single workstation GPU under a
fixed serving stack.

\subsection{General-Purpose LLM Benchmarking}
\label{sec:related-benchmarking}

Our evaluation protocol is also informed by broader LLM benchmarking
practice. MMLU~\cite{hendrycks2021mmlu} evaluates language models across
many knowledge domains; HELM~\cite{liang2022helm} advocates holistic
evaluation across multiple dimensions; and
SWE-bench~\cite{jimenez2023swebench} evaluates software-engineering
capability using execution-grounded tasks.

These benchmarks illustrate the importance of evaluating model behavior
along multiple dimensions rather than relying on a single aggregate
score. TriCalRAG applies the same principle to AIOps by reporting
detection metrics together with prediction-balance and deployment
diagnostics.

\subsection{Calibration of In-Context Learning}
\label{sec:related-calibration}

Large language models can exhibit systematic label bias under zero-shot
and few-shot prompting. Zhao et al.~\cite{zhao2021calibrate} show that
predictions may be influenced by label frequency, prompt ordering, and
other contextual effects, and propose \emph{Contextual Calibration} (CC)
to estimate and correct such bias using content-free inputs.

Zhou et al.~\cite{zhou2024batch} propose \emph{Batch Calibration} (BC),
which estimates label bias from the average prediction over an unlabeled
batch and therefore does not require auxiliary content-free examples.
Both methods were developed primarily for general NLP classification
settings.

TriCalRAG evaluates CC and BC in a separate next-token
log-probability experiment to test whether the extreme prediction
imbalance observed in our main benchmark can be explained by established
in-context label-bias mechanisms. Importantly, our
predicted-positive-rate diagnostic is not itself a measure of
probabilistic calibration. The calibration methods therefore serve as
mechanism-checking baselines rather than alternative definitions of the
main evaluation metric.

Section~\ref{sec:calibration-rq2} compares these methods with the raw
log-probability decision rule and examines whether they reproduce the
reduction in prediction imbalance observed under RAG.


Table~\ref{tab:related-work} summarizes the representative approaches
discussed above. Existing work separately addresses log anomaly
detection, LLM-based diagnosis, retrieval augmentation, model serving,
and in-context calibration. TriCalRAG brings these dimensions together
in a controlled benchmark focused on single-GPU on-premise deployment,
prompting strategy, prediction balance, and inference efficiency.

\section{TriCalRAG Benchmark Design}
\label{sec:benchmark}

\begin{figure}[!htbp]
    \centering
    \includegraphics[width=0.98\linewidth]{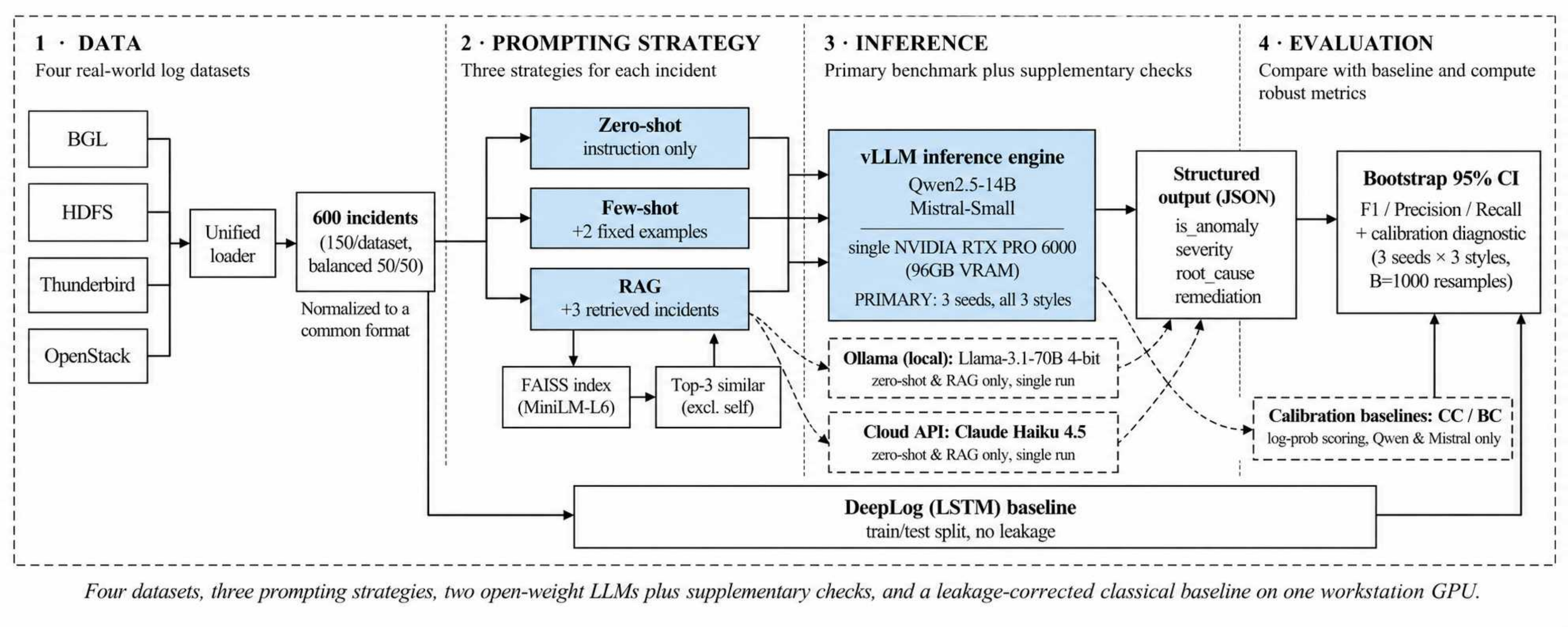}
    \caption{TriCalRAG evaluation architecture.}
    \label{fig:tricalrag}
\end{figure}

This section presents the design of TriCalRAG, a benchmark for systematically evaluating LLM-based incident analysis in practical on-premise AIOps settings. Fig.~\ref{fig:tricalrag} illustrates the overall TriCalRAG evaluation architecture. We first define the benchmark task and datasets, followed by incident construction, prompting strategies, the retrieval pipeline, evaluated systems, and the benchmark protocol. 


\subsection{Task Definition}
\label{sec:task-definition}

Each benchmark instance is an \emph{incident}: a short window of raw
operational log lines drawn from one of the four datasets. Given an
incident, an LLM must produce a structured JSON object containing a
binary \texttt{is\_anomaly} judgment, a \texttt{severity} level, a
one-sentence \texttt{root\_cause} explanation, and a one-sentence
\texttt{remediation} suggestion.

The binary anomaly decision is evaluated quantitatively. The generated
root-cause and remediation fields are retained as part of the RCA output
but are not automatically scored in the current study; the implications
of this evaluation scope are discussed in
Section~\ref{sec:limitations}.

\subsection{Datasets}
\label{sec:datasets}
We use four public LogHub~\cite{loghub} datasets representing diverse operational environments, as summarized in Fig.~\ref{fig:scope}, which illustrates the overall benchmark scope.

\begin{itemize}
    \item \textbf{BGL}: BlueGene/L supercomputer logs with per-line
    alert-category labels.

    \item \textbf{HDFS}: distributed-filesystem logs with anomaly labels
    defined at the block level.

    \item \textbf{Thunderbird}: large-scale cluster logs with per-line
    labels.

    \item \textbf{OpenStack}: cloud-infrastructure logs distributed as
    separate normal and injected-anomaly files.
\end{itemize}

From each dataset we construct 150 incidents, balanced equally between
anomalous and normal examples, yielding 600 incidents per evaluation
run. Because the raw Thunderbird log is approximately 31.7\,GB, we
sample from its first two million lines.

\subsection{Incident Construction}
\label{sec:incident-construction}

Each incident is represented as a window of five consecutive raw log
lines. The same fixed-window representation is used across all four
datasets to provide a uniform model input interface.

This representation aligns naturally with line-labeled datasets such
as BGL and Thunderbird. HDFS, however, is labeled at the block level,
and log entries belonging to the same block may occur non-contiguously.
Consequently, a five-line window can split evidence associated with a
single HDFS block. We retain the common incident representation for the
current benchmark to preserve protocol consistency and discuss this
evaluation-unit mismatch explicitly in
Section~\ref{sec:limitations}.

\subsection{Prompting Strategies}
\label{sec:prompting}

All LLM configurations receive the same core task instruction. The three prompting strategies differ only in the additional context provided to the model.

\begin{itemize}
    \item \textbf{Zero-shot.} The model receives the task instruction and the target incident only.
    \item \textbf{Few-shot.} Two fixed worked examples are prepended to the prompt. These examples remain identical across all queries.
    \item \textbf{RAG.} The prompt is augmented with three semantically similar labeled historical incidents retrieved specifically for the target query.
\end{itemize}

This controlled design isolates prompt context as the primary variable while keeping the output schema and underlying anomaly-detection task constant.

\subsection{Retrieval Pipeline}
\label{sec:retrieval}

For the RAG condition, incidents are embedded using
\texttt{all-MiniLM-L6-v2}, a Sentence-BERT model
~\cite{reimers2019sbert}. The embeddings are indexed using
FAISS~\cite{johnson2019faiss}, and cosine similarity is used to retrieve
the three most similar labeled incidents.

The retrieved incidents and their ground-truth outcomes are supplied as
precedent in the prompt. The query incident itself is excluded from the
retrieval results to prevent direct self-retrieval.

Query exclusion prevents the most direct form of label leakage, although
it does not by itself establish the absence of near-duplicate or
corpus-level overlap. We therefore treat broader retrieval leakage as a
validity consideration rather than claiming complete leakage elimination.

\subsection{Systems Evaluated}
\label{sec:systems}

The primary benchmark evaluates two open-weight instruction-tuned LLMs:
Qwen2.5-14B-Instruct~\cite{qwen2024technical} and
Mistral-Small-Instruct-2409 (22B)~\cite{mistralsmall2409}.
Both are served locally in bfloat16 through
vLLM~\cite{vllm2023} on the same workstation GPU.

We additionally evaluate a DeepLog-style LSTM
baseline~\cite{deeplog2017} trained on normal sequences and assessed
through a separate held-out protocol. Because DeepLog requires explicit
training and uses a different evaluation split, it is treated as a
classical reference baseline rather than as another member of the
primary LLM protocol.

Two supplementary model checks extend the evaluation beyond the primary
models: a 4-bit Llama-3.1-70B-Instruct model served locally through
Ollama and Claude Haiku 4.5 accessed through the Anthropic API. These
supplementary experiments use single-run protocols and are therefore
reported separately from the repeated primary benchmark.

\subsection{Benchmark Protocol Summary}
\label{sec:benchmark-summary}

\begin{figure}[t]
\centering

\begin{minipage}[t]{0.47\columnwidth}
\centering

\vspace{0pt}

\includegraphics[
    width=\linewidth
]{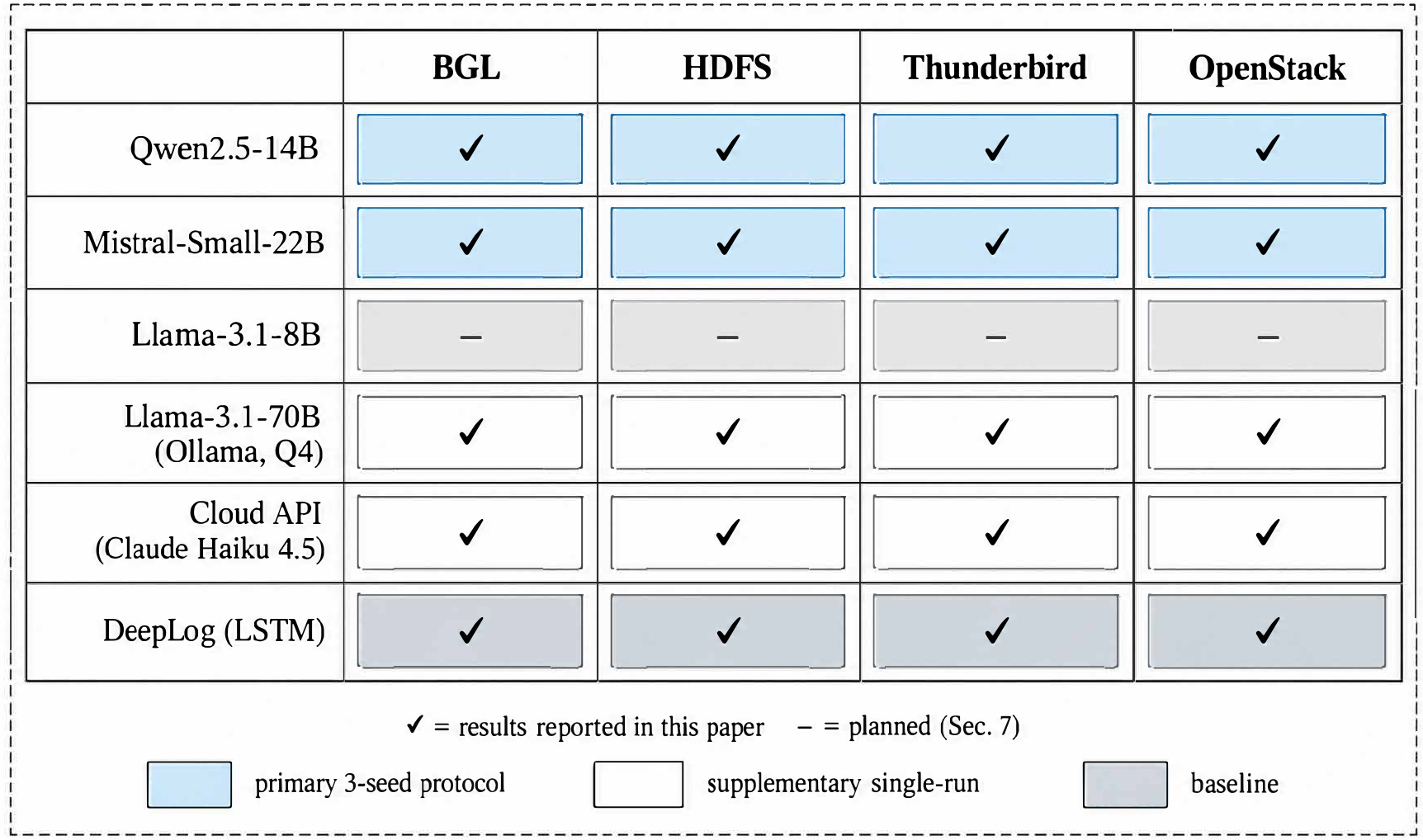}

\captionof{figure}{%
Benchmark scope. Qwen2.5-14B and Mistral-Small-22B form
the repeated primary LLM protocol, while DeepLog provides the
classical baseline. Llama-3.1-70B through Ollama and Claude
Haiku 4.5 through the Anthropic API are reported as
supplementary single-run evaluations
(Sections~\ref{sec:llama70b} and~\ref{sec:cloud}).
}
\label{fig:scope}

\end{minipage}
\hfill
\begin{minipage}[t]{0.47\columnwidth}
\centering

\vspace{0pt}

\captionof{table}{TriCalRAG benchmark construction and evaluation scope.}
\label{tab:benchmark-protocol}

{\scriptsize
\renewcommand{\arraystretch}{1.08}
\setlength{\tabcolsep}{2.5pt}

\begin{tabular}{
    @{}
    >{\raggedright\arraybackslash}p{0.37\linewidth}
    >{\raggedright\arraybackslash}p{0.57\linewidth}
    @{}
}
\toprule
\textbf{Property} & \textbf{TriCalRAG setting} \\
\midrule

Datasets
& BGL, HDFS, Thunderbird, OpenStack \\

Incidents / dataset
& 150 \\

Incidents / run
& 600 \\

Class distribution
& 50\% anomalous / 50\% normal \\

Incident unit
& Five consecutive log lines \\

Primary LLMs
& Qwen2.5-14B, Mistral-Small-22B \\

Prompt strategies
& Zero-shot, Few-shot, RAG \\

Embedding model
& all-MiniLM-L6-v2 \\

Vector index
& FAISS \\

RAG context
& Top-3 labeled incidents \\

Primary repetitions
& 3 data-sampling seeds \\

Local serving
& vLLM, RTX PRO 6000 (96\,GB) \\

Primary metric
& F1 with bootstrap 95\% CI \\

Balance diagnostic
& Predicted-positive rate \\

Classical baseline
& DeepLog \\

Supplementary
& Llama-3.1-70B, Claude Haiku 4.5 \\

\bottomrule
\end{tabular}
}

\end{minipage}

\end{figure}

Table~\ref{tab:benchmark-protocol} summarizes the principal design
choices of TriCalRAG. The primary benchmark consists of four datasets,
two locally served LLMs, three prompting strategies, and three
data-sampling seeds. Execution and statistical details are provided in
Section~\ref{sec:expsetup}.

\section{Evaluation Framework}
\label{sec:expsetup}

This section describes the research questions, hardware environment,
primary experimental protocol, evaluation metrics, statistical
methodology, supplementary evaluations, deployment ablations, and
reproducibility scope of TriCalRAG. We separate the repeated primary
benchmark from single-run supplementary experiments so that experiments
conducted under different protocols are not assigned the same level of
statistical confidence.

\subsection{Research Questions}
\label{sec:rqs}

Our evaluation is organized around four research questions.

\textbf{RQ1: Detection effectiveness.}
How do zero-shot, few-shot, and retrieval-augmented generation (RAG)
prompting affect anomaly-detection performance across models and
datasets?

\textbf{RQ2: Prediction balance.}
To what extent can aggregate F1 obscure degenerate prediction behavior,
and does RAG reduce extreme predicted-positive rates relative to
zero-shot and few-shot prompting?

\textbf{RQ3: Mechanism and generalization.}
Can established in-context calibration methods explain the behavior
observed under RAG, and does the observed prediction-balance effect
extend to a larger open-weight model and a cloud-served model?

\textbf{RQ4: Deployment efficiency.}
What throughput, latency, memory, quantization, and API-cost trade-offs
arise when deploying TriCalRAG locally on a single workstation GPU
versus using a cloud-hosted model?

\subsection{Systems Evaluated}
\label{sec:systems-evaluated}

The primary benchmark evaluates two open-weight instruction-tuned
models: Qwen2.5-14B-Instruct~\cite{qwen2024technical} and
Mistral-Small-Instruct-2409 (22B)~\cite{mistralsmall2409}.
Both are served locally through vLLM~\cite{vllm2023} in bfloat16
precision.

The primary models are evaluated under three prompting strategies:
zero-shot, few-shot, and RAG. As described in
Section~\ref{sec:retrieval}, the RAG pipeline embeds incidents using
\texttt{all-MiniLM-L6-v2}, indexes them using
FAISS~\cite{johnson2019faiss}, and retrieves the top-3 most similar
labeled incidents while excluding the query itself.

We additionally evaluate a DeepLog~\cite{deeplog2017} LSTM baseline
through a separate held-out protocol. Because the DeepLog baseline
requires model training and is evaluated on a smaller held-out set, its
results are treated as a complementary classical baseline rather than as
a directly paired run of the primary LLM protocol.

\subsection{Datasets and Evaluation Unit}
\label{sec:eval-datasets}

TriCalRAG evaluates four public LogHub~\cite{loghub} datasets:
BGL, HDFS, Thunderbird, and OpenStack. Each evaluation run contains
150 incidents from each dataset, balanced 50/50 between anomalous and
normal examples, for a total of 600 incidents per run.

Each incident is represented as a five-line log window and is presented
to the model with an instruction requiring a structured JSON response
containing the binary anomaly decision together with severity,
root-cause, and remediation fields. Quantitative evaluation in the
current study focuses on the binary \texttt{is\_anomaly} decision;
the semantic quality of the generated root-cause and remediation text
is not automatically scored.

Additional details on dataset construction and the HDFS evaluation-unit
mismatch are provided in Section~\ref{sec:incident-construction}.

\subsection{Hardware and Local Serving}
\label{sec:hardware}

All primary local-model experiments were executed on a single
workstation equipped with an NVIDIA RTX PRO 6000 GPU with 96\,GB of
VRAM. Qwen2.5-14B and Mistral-Small-22B were served through
vLLM~\cite{vllm2023}.

This single-workstation setting is intentional: TriCalRAG evaluates
whether useful LLM-based log analysis can be performed without
requiring a multi-GPU datacenter-class serving environment. In addition
to predictive metrics, we therefore record deployment-oriented
measurements including throughput and GPU memory consumption.

\subsection{Primary Experimental Protocol}
\label{sec:primary-protocol}

The primary experiment evaluates the Cartesian product of two models,
four datasets, and three prompting strategies:

\[
2\ \text{models}
\times 4\ \text{datasets}
\times 3\ \text{prompt styles}.
\]

Each model--dataset--prompt configuration is evaluated across three
independent data-sampling seeds. With 150 incidents per dataset, the
primary benchmark yields 10,800 incident-level model responses across
the repeated evaluation.

The three prompting conditions are:

\begin{itemize}

    \item \textbf{Zero-shot:}
    the model receives the task instruction and the target incident
    only.

    \item \textbf{Few-shot:}
    two fixed worked examples are prepended to the task instruction and
    reused across queries.

    \item \textbf{RAG:}
    the task instruction is augmented with the top-3 semantically
    similar labeled historical incidents retrieved through the
    MiniLM/FAISS pipeline.

\end{itemize}

Prompt style is therefore the primary controlled variable in the main
benchmark.

\subsection{Evaluation Metrics}
\label{sec:metrics}

We report classification effectiveness using precision, recall, and
F1-score. Because the benchmark is constructed with a true anomaly
prevalence of 0.50, we additionally report the
\emph{predicted-positive rate} (PPR), defined as the fraction of
incidents that a configuration labels anomalous.

PPR is used as a prediction-balance diagnostic rather than as a measure
of probabilistic calibration. A configuration that labels nearly every
incident anomalous can obtain high recall and a moderate-looking F1
while providing little discriminative value.

Following the predefined diagnostic used throughout the study, we flag
a configuration as \emph{degenerate} when

\[
\mathrm{PPR} < 0.15
\qquad \text{or} \qquad
\mathrm{PPR} > 0.85.
\]

The true 0.50 anomaly rate is shown in the result figures as a reference
point, but proximity to 0.50 alone does not establish probabilistic
calibration or classification correctness.

For deployment efficiency, we additionally report model throughput in
tokens per second, GPU memory consumption where applicable, and latency
for the deployment ablations.

\subsection{Statistical Methodology}
\label{sec:statistics}

Each primary configuration
(model $\times$ dataset $\times$ prompt style) is evaluated across
three random data-sampling seeds. We report
bootstrap~\cite{efron1979bootstrap} 95\% confidence intervals on
F1 scores using 1,000 bootstrap resamples.

The repeated three-seed protocol applies to the primary Qwen2.5-14B
and Mistral-Small-22B experiments. Supplementary experiments described
below are reported separately and should not be interpreted with the
same statistical confidence as the repeated primary benchmark.

\subsection{Calibration-Correction Baselines}
\label{sec:calibration-baselines}

To test whether the extreme prediction imbalance observed in the
generated JSON decisions can be explained by established in-context
label-bias mechanisms, we conduct a separate scoring experiment using
Contextual Calibration (CC)~\cite{zhao2021calibrate} and
Batch Calibration (BC)~\cite{zhou2024batch}.

This experiment reads the model's next-token log-probability for the
binary anomaly label and applies CC or BC to that score. This
log-probability decision path is distinct from the parsed structured
JSON outputs used in the main benchmark. Consequently, the resulting
degeneracy counts are interpreted as an internally consistent mechanism
check and are not treated as numerically interchangeable with the
primary F1 results.

\subsection{Supplementary Model Evaluations}
\label{sec:supplementary-protocol}

We conduct two supplementary checks outside the primary three-seed
protocol.

First, Llama-3.1-70B-Instruct is evaluated in 4-bit quantized form
through Ollama under zero-shot and RAG prompting. This experiment is
reported as a single-run supplementary evaluation and is used to
examine whether the observed dataset-dependent prediction imbalance
extends to a substantially larger open-weight model.

Second, Claude Haiku 4.5 is evaluated through the Anthropic API on the
same 600-incident benchmark under zero-shot and RAG prompting. This is
also a single-run supplementary experiment and is used to examine the
same behavior under a cloud-served model and to ground the deployment
cost discussion.

Because these experiments use a different repetition protocol from the
primary benchmark, their results are presented as supplementary
evidence rather than as additional members of the primary statistical
comparison.

\subsection{Deployment Ablations}
\label{sec:deployment-ablations}

We evaluate two deployment-oriented factors on the local setup.

\textbf{Batch-size scaling.}
Using Qwen2.5-14B on the RTX PRO 6000, we evaluate batch sizes from
1 to 128 to characterize the throughput available from a single GPU
under increasing concurrency.

\textbf{Quantization.}
We compare the bfloat16 Qwen2.5-14B configuration against an
AWQ~\cite{lin2024awq} 4-bit variant on the same workload, reporting
F1 and latency to quantify the accuracy--efficiency trade-off.

These ablations characterize deployment behavior rather than modify
the primary model-comparison protocol.

\subsection{Reproducibility Scope}
\label{sec:reproducibility}

The primary benchmark consists of 72 dataset-level repeated
configurations when the three data-sampling seeds are included:
\[
2\ \text{models}
\times 4\ \text{datasets}
\times 3\ \text{prompt styles}
\times 3\ \text{seeds}
= 72.
\]

At 150 incidents per dataset, this corresponds to 10,800
incident-level model responses. The benchmark harness separates dataset
loading, prompt construction, retrieval, inference, result scoring, and
figure generation so that each stage can be reproduced independently.

The single-run Llama-3.1-70B and Claude Haiku 4.5 experiments, the
DeepLog baseline, and the calibration-correction experiment are reported
separately because they do not share the complete primary three-seed
protocol.

\section{Experimental Results}
\label{sec:results}

This section presents the experimental results organized around four research questions: RQ1. evaluates detection effectiveness, RQ2. examines prediction balance, RQ3. investigates the underlying mechanisms and generalization of the observed behavior, and RQ4. assesses deployment efficiency under practical on-premise constraints.

\subsection{RQ1: Detection Effectiveness}
\label{sec:rq1}

Table~\ref{tab:main-results} reports macro-averaged results across the
four datasets. Mistral-Small achieves the higher mean F1
(0.644 vs.\ 0.560), while Qwen2.5-14B exhibits fewer extreme
prediction-balance cases across the evaluated dataset--prompt
combinations and approximately twice the inference throughput.
JSON parse failures were rare, affecting only 0.2\% of the
10{,}800 model responses.

\begin{table}[t]
\centering
\caption{Macro-averaged results across all 4 datasets
(mean over 3 seeds, 2 models, and 3 prompt styles).}
\label{tab:main-results}
\begin{tabular}{lcccc}
\toprule
Model & Mean F1 & Pred.\ Pos.\ Rate & Tok/s & VRAM (GB) \\
\midrule
Mistral-Small & 0.644 & 0.713 & 365.9 & 86.0 \\
Qwen2.5-14B   & 0.560 & 0.485 & 713.1 & 86.6 \\
\bottomrule
\end{tabular}
\end{table}

\subsubsection{Per-Dataset Breakdown}

\begin{figure*}[t]
\centering
\includegraphics[
    width=0.96\textwidth
]{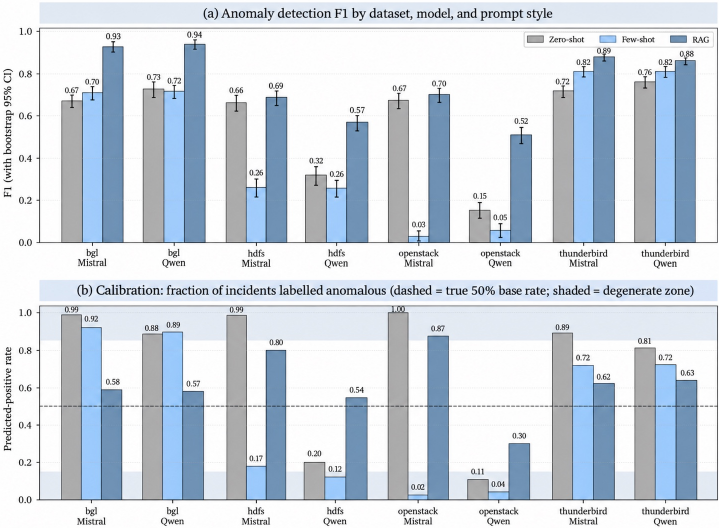}

\vspace{-1mm}

\caption{Benchmark results across four datasets, two models, and three prompting strategies, (a) Anomaly-detection F1 and  (b) Predicted-positive rate.}
\label{fig:results}
\end{figure*}

Performance varies substantially by dataset and prompt style. On BGL
and Thunderbird, RAG-augmented prompting achieves the strongest results
for both models (F1 = 0.92--0.94 on BGL, F1 = 0.88--0.89 on
Thunderbird), compared with F1 = 0.51--0.73 under zero-shot prompting
on the same datasets. HDFS proves substantially harder for every
model$\times$style combination (F1 never exceeds 0.687), which we
attribute to a windowing mismatch discussed in
Section~\ref{sec:limitations}: HDFS anomaly labels are assigned per
block ID, but our fixed 5-line contiguous windowing can split a single
block's related log lines across multiple windows, misaligning the
evaluation unit with the true anomaly unit.

\subsection{RQ2: Prediction Balance}
\label{sec:calibration}

Beyond raw F1, we report each configuration's predicted-positive rate
(PPR), the fraction of incidents a model labels ``anomaly.'' Because
the benchmark has a balanced 50\% true anomaly rate, PPR provides a
simple diagnostic for extreme class-distribution behavior. It is not a
measure of probabilistic calibration.

A model that predicts ``anomaly'' indiscriminately can achieve high
recall and a deceptively reasonable F1 while providing little
discriminative value. We flag configurations with
$\mathrm{PPR}>0.85$ or $\mathrm{PPR}<0.15$ as \emph{degenerate}.
Figure~\ref{fig:results}(b) visualizes this behavior.

Zero-shot prompting drives both models toward degenerate behavior on
3 of 4 datasets. Mistral-Small's zero-shot PPR reaches 0.993 on BGL,
0.987 on HDFS, and 1.000 on OpenStack, while its zero-shot accuracy on
these datasets (0.500--0.507) is barely above chance despite F1 scores
of 0.67--0.72. Qwen2.5-14B shows the same pattern, though somewhat less
severely.

RAG substantially reduces this behavior: 7 of the 8 RAG configurations
remain within the predefined PPR range, compared with only 2 of 8
zero-shot configurations. Few-shot prompting shows an opposite failure
mode on HDFS and OpenStack, where both models under-predict anomalies
(PPR as low as 0.016), suggesting that fixed examples can also induce
strong class-distribution bias.

\subsection{RQ3: Mechanism and Generalization}
\label{sec:rq3}

\subsubsection{Calibration-Correction Baselines}
\label{sec:calibration-rq2}

To test whether the degenerate behavior above is attributable to the
general in-context label bias described by Zhao et
al.~\cite{zhao2021calibrate} and Zhou et al.~\cite{zhou2024batch}, we
re-score all 24 configurations using a log-probability-based decision
rule, reading
$P(\texttt{is\_anomaly}=\text{true})$ directly from next-token
log-probabilities rather than parsing generated JSON text, and apply
Contextual Calibration (CC) and Batch Calibration (BC).

This scoring path is separate from the text-parsed results reported
above; see Section~\ref{sec:limitations} for how the two should and
should not be compared.

Table~\ref{tab:calib-baselines} summarizes the results. Under the raw
log-probability score, 11 of 24 configurations are degenerate,
concentrated in zero-shot (5/8) and few-shot (6/8), while RAG remains
at 0/8. CC reduces degeneracy under zero-shot (5$\to$3) and few-shot
(6$\to$5) but introduces degeneracy in two RAG configurations.
BC provides no net improvement (12/24 degenerate versus 11/24 raw).

These results indicate that RAG's reduction in prediction imbalance is
not reproduced by correcting a static label-frequency prior. Unlike CC
and BC, which modify the output distribution independently of the
query, RAG supplies query-specific labeled precedent.

\subsubsection{Generalization to Llama-3.1-70B}
\label{sec:llama70b}

To test whether this behavior is specific to Qwen2.5-14B and
Mistral-Small, we repeat the zero-shot/RAG comparison with
Llama-3.1-70B-Instruct (4-bit quantized, served locally via
Ollama\footnote{\url{https://ollama.com}}) as a single-run
supplementary check outside the primary three-seed protocol.

Under zero-shot prompting, BGL and Thunderbird show over-prediction
(PPR = 0.913 and 0.820), while HDFS and OpenStack show under-prediction
(0.393 and 0.160). Under RAG, the corresponding rates become 0.600,
0.680, 0.533, and 0.407, respectively, and F1 improves on every
dataset, most substantially on OpenStack (0.222 to 0.618).

\begin{table*}[!t]
\centering

\begin{minipage}[t]{0.44\textwidth}
\centering
\caption{Degenerate configurations by decision rule and prompt style
(of 8 each; 24 total).}
\label{tab:calib-baselines}

\vspace{2mm}

\begin{tabular}{lccc}
\toprule
Prompt style & Raw & CC & BC \\
\midrule
Zero-shot & 5/8 & 3/8 & 6/8 \\
Few-shot  & 6/8 & 5/8 & 6/8 \\
RAG       & 0/8 & 2/8 & 0/8 \\
\midrule
Total (of 24) & 11 & 10 & 12 \\
\bottomrule
\end{tabular}
\\ \smallskip
{\footnotesize 
CC: Contextual Calibration, BC: Batch Calibration}
\end{minipage}
\hfill
\begin{minipage}[t]{0.54\textwidth}
\centering
\caption{Zero-shot vs.\ RAG, Llama-3.1-70B
(single run, 4-bit, via Ollama).}
\label{tab:llama70b}

\vspace{2mm}

\resizebox{\linewidth}{!}{%
\begin{tabular}{lcccc}
\toprule
Dataset &
\multicolumn{2}{c}{Pred.\ Pos.\ Rate} &
\multicolumn{2}{c}{F1} \\
& Zero-shot & RAG & Zero-shot & RAG \\
\midrule
BGL         & 0.913 & 0.600 & 0.708 & 0.909 \\
HDFS        & 0.393 & 0.533 & 0.463 & 0.568 \\
OpenStack   & 0.160 & 0.407 & 0.222 & 0.618 \\
Thunderbird & 0.820 & 0.680 & 0.758 & 0.847 \\
\bottomrule
\end{tabular}%
}
\end{minipage}

\end{table*}

\begin{figure}[t]
\centering
\includegraphics[width=\columnwidth]{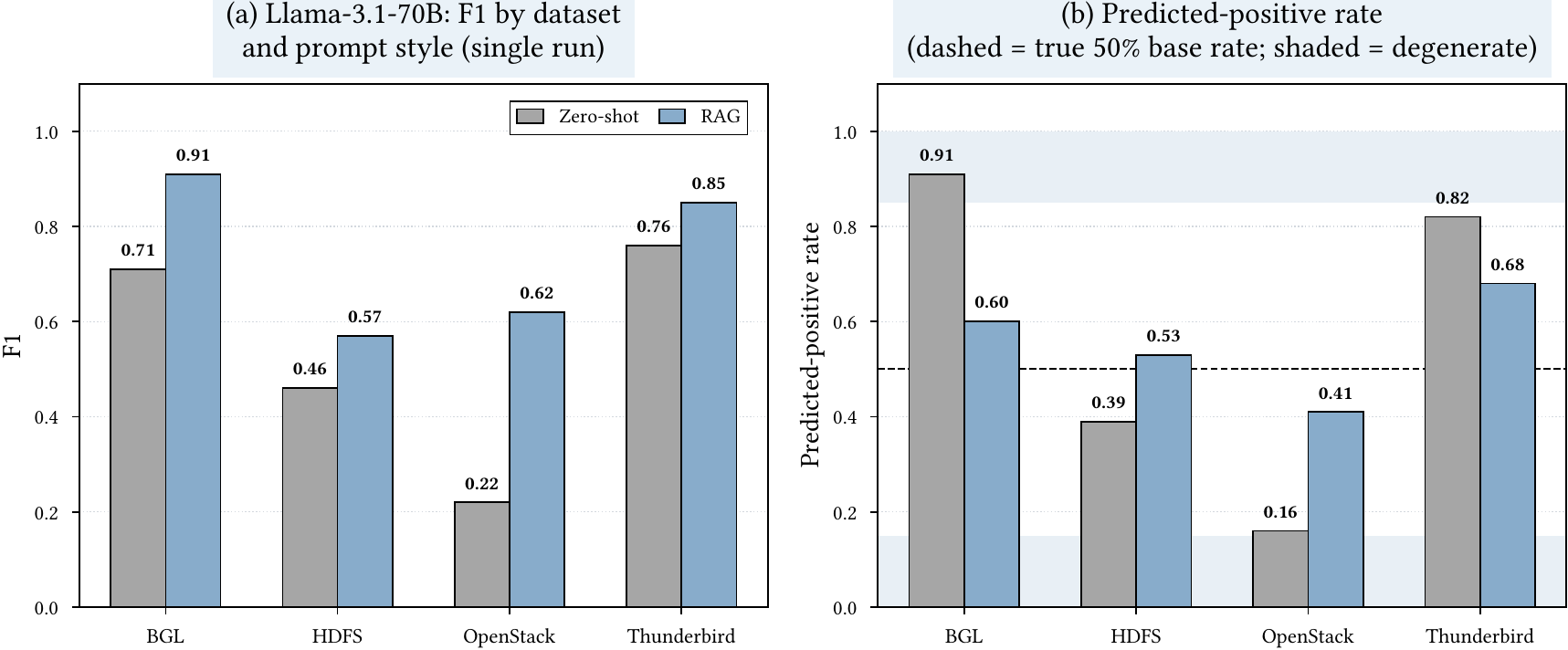}
\caption{Llama-3.1-70B (single run, 4-bit, via Ollama):
(a) F1 by dataset under zero-shot and RAG prompting;
(b) predicted-positive rate, with the degenerate regions shaded and the
true 50\% base rate dashed. Zero-shot prediction imbalance occurs in
both directions across datasets, while RAG generally moves predictions
toward the base rate.}
\label{fig:llama70b}
\end{figure}

\subsection{RQ4: Deployment Efficiency}
\label{sec:rq4}

\subsubsection{Batch-Size Scaling}

Table~\ref{tab:batch-sweep} reports Qwen2.5-14B throughput across batch
sizes on the RTX PRO 6000. Throughput increases from 48.3 tokens/s at
batch size 1 to 1991.0 tokens/s at batch size 128, approximately a
41$\times$ improvement, while batch elapsed time increases from 1.39s
to 4.14s.

\begin{table*}[!t]
\centering

\begin{minipage}[t]{0.48\textwidth}
\centering
\caption{Batch-size scaling, Qwen2.5-14B on one RTX PRO 6000.}
\label{tab:batch-sweep}

\begin{tabular}{ccc}
\toprule
Batch size & Elapsed (s) & Throughput (tok/s) \\
\midrule
1   & 1.39 & 48.3 \\
8   & 2.09 & 272.8 \\
32  & 2.95 & 723.9 \\
64  & 3.29 & 1296.4 \\
128 & 4.14 & 1991.0 \\
\bottomrule
\end{tabular}
\end{minipage}
\hfill
\begin{minipage}[t]{0.40\textwidth}
\centering
\caption{Quantization comparison, Qwen2.5-14B.}
\label{tab:quantization}

\begin{tabular}{lcc}
\toprule
Variant & F1 & Latency (s) \\
\midrule
bfloat16  & 0.730 & 5.06 \\
AWQ 4-bit & 0.739 & 4.04 \\
\bottomrule
\end{tabular}
\end{minipage}

\end{table*}

\subsubsection{Quantization Impact}

We compare Qwen2.5-14B in bfloat16 against its
AWQ~\cite{lin2024awq} 4-bit variant on an identical incident subset.
The quantized model achieves F1 = 0.739 versus 0.730 for bfloat16,
while latency falls from 5.06s to 4.04s, a 20\% reduction. The observed
F1 difference should not be interpreted as a statistically established
improvement; the practical result is that quantization reduces latency
without measurable degradation on the tested workload.

\section{Discussion}
\label{sec:discussion}

Our results suggest that the choice between Mistral-Small and Qwen2.5-14B
for on-premise RCA is not simply a matter of picking the higher-F1 model.
Mistral-Small's F1 advantage is concentrated in configurations where its
prediction balance is comparatively stable (RAG and few-shot on
BGL/Thunderbird); in zero-shot settings, its apparent competence is
substantially influenced by near-constant positive prediction. A deployment
that cannot guarantee retrieval infrastructure or curated few-shot examples
at inference time---for instance, a cold-start incident with no similar
historical precedent to retrieve---may therefore experience substantially
less reliable class balance under zero-shot prompting. This is a practically
important distinction that a single aggregate F1 number obscures, and we
recommend that future LLM-for-RCA evaluations report predicted-positive rate
(or an equivalent prediction-balance diagnostic) alongside F1 as standard
practice.

The strength of RAG in this benchmark is consistent with the intuition
that retrieved historical incidents provide the model with concrete,
task-relevant priors on what ``anomaly'' looks like in a given log format.
This can act as a form of query-specific in-context guidance that generic
few-shot examples (written once and reused across all queries) cannot provide,
because the retrieved context changes with each incident's nearest neighbors.
We treat predicted-positive rate as a diagnostic of class balance rather than
as proof of probabilistic calibration.

\subsection{Comparison to a Classical Baseline}

We compare against DeepLog~\cite{deeplog2017}, a canonical LSTM-based log
anomaly detector, re-implemented with a log-key next-token prediction
objective trained exclusively on normal sequences. To ensure a fair
comparison, we evaluate DeepLog on a held-out set disjoint from its
training data (an 80/20 split of normal sequences, with the held-out 20\%
combined with a randomly sampled equal number of anomalous sequences to
match the LLM benchmark's balanced 50/50 evaluation protocol)---an
important methodological correction, since evaluating on training-set
normal sequences (data leakage) initially produced an inflated F1 of 0.913
that did not reflect genuine generalization.

Under this corrected protocol, DeepLog achieves F1 = 0.698
(Precision = 0.584, Recall = 0.867), numerically above the reported
macro-averaged F1 of both Qwen2.5-14B (0.560) and Mistral-Small (0.644).
However, this should not be interpreted as a direct superiority result
because DeepLog is evaluated on a substantially smaller held-out set under
a different training and evaluation protocol. Its predicted-positive rate is
0.742---meaningfully elevated relative to the true 50\% base rate, though
below our degenerate threshold. This indicates a tendency toward
over-predicting anomalies at this scale of training data (240 training
sequences per dataset). Unlike the LLM-based approaches, DeepLog produces
no natural-language root-cause explanation, only a binary anomaly flag.
For practitioners who need a system to explain \emph{why} an incident is
anomalous rather than only flag \emph{that} it is, this is an important
functional distinction independent of raw F1.

We note that this comparison uses a substantially smaller held-out evaluation
set (120 incidents total) than the LLM benchmark's per-run evaluation
(600 incidents), since DeepLog's training requirement consumes most of
the available normal-labeled sequences. This asymmetry is discussed
further in Section~\ref{sec:limitations}.

\subsection{Comparison to a Cloud API}
\label{sec:cloud}

\begin{figure}[t]
\centering
\includegraphics[width=\columnwidth]{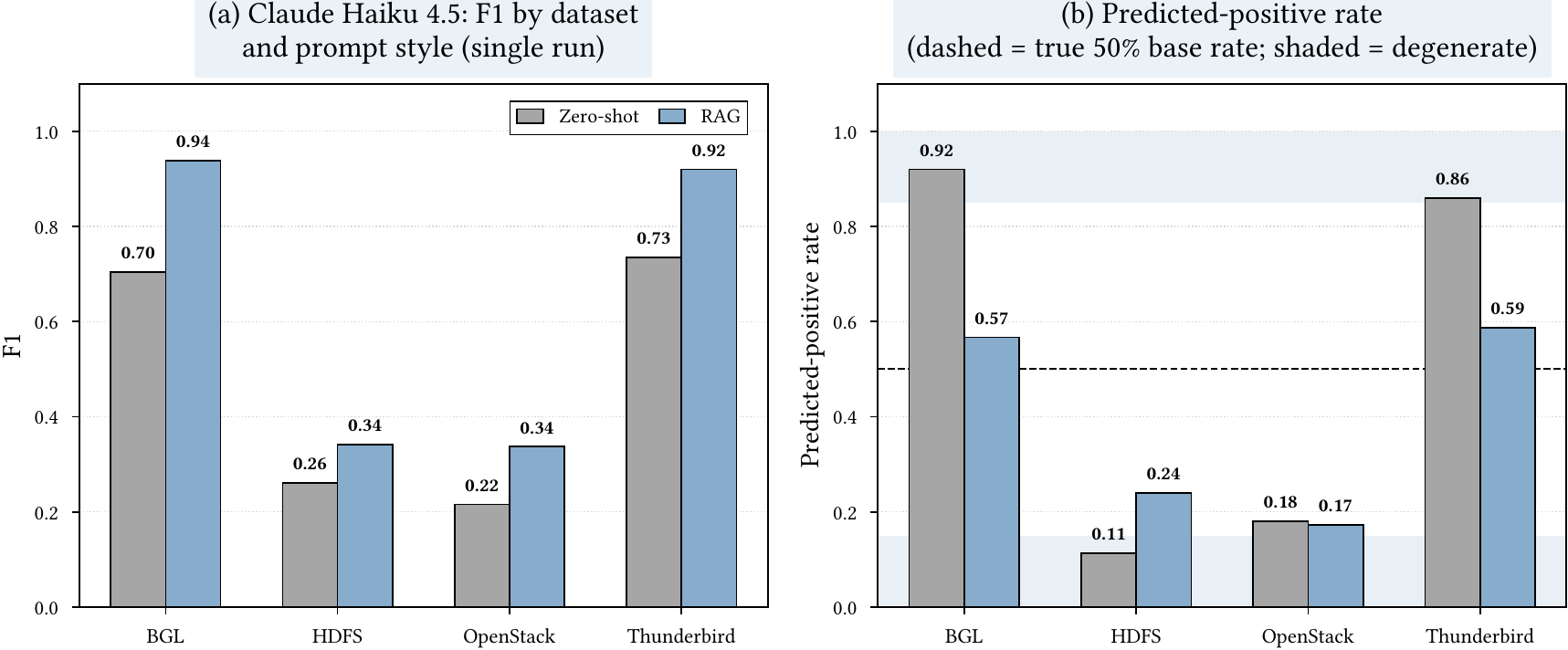}
\caption{Claude Haiku 4.5 (single run, native API): (a) F1 by dataset
under zero-shot and RAG prompting; (b) predicted-positive rate, with the
degenerate threshold shaded and the true 50\% base rate dashed. As with
Llama-3.1-70B (Fig.~\ref{fig:llama70b}), aggregate predicted-positive
rates can conceal per-dataset prediction imbalance that appears only when
results are disaggregated.}
\label{fig:cloud}
\end{figure}

As a preliminary check on the cost argument motivating this benchmark, we
additionally evaluate Claude Haiku 4.5 via Anthropic's cloud API on the
same 600 incidents under zero-shot and RAG prompting (single run, outside
the primary 3-seed protocol). Aggregated across all four datasets, Claude
Haiku 4.5 achieves F1 = 0.566 (zero-shot) and F1 = 0.695 (RAG) with an
overall predicted-positive rate of 0.518 and 0.392, respectively. These
aggregate values obscure substantial per-dataset variation. Breaking the
results down by dataset (Table~\ref{tab:cloud} and Fig.~\ref{fig:cloud})
shows that under zero-shot prompting, three of four datasets cross the
predefined degenerate thresholds: BGL = 0.920 and Thunderbird = 0.860
over-predict, while HDFS = 0.113 under-predicts. OpenStack = 0.180 remains
inside the non-degenerate range. Under RAG, those three threshold failures
are removed (BGL = 0.567, Thunderbird = 0.587, HDFS = 0.240), while
OpenStack moves from 0.180 to 0.173. F1 improves on every dataset, most
sharply on BGL (0.704 to 0.938) and Thunderbird (0.735 to 0.920).

\begin{table}[t]
\centering
\caption{Zero-shot vs.\ RAG, Claude Haiku 4.5 (single run, native API).}
\label{tab:cloud}
\begin{tabular}{lcccc}
\toprule
Dataset & \multicolumn{2}{c}{Pred.\ Pos.\ Rate} & \multicolumn{2}{c}{F1} \\
        & Zero-shot & RAG & Zero-shot & RAG \\
\midrule
BGL         & 0.920 & 0.567 & 0.704 & 0.938 \\
HDFS        & 0.113 & 0.240 & 0.261 & 0.342 \\
OpenStack   & 0.180 & 0.173 & 0.216 & 0.337 \\
Thunderbird & 0.860 & 0.587 & 0.735 & 0.920 \\
\bottomrule
\end{tabular}
\end{table}

The \emph{direction} of zero-shot prediction imbalance by dataset matches
the pattern found independently in Llama-3.1-70B
(Section~\ref{sec:llama70b}, Fig.~\ref{fig:llama70b}): both models
over-predict on BGL and Thunderbird and under-predict on HDFS and
OpenStack, despite being different model families served through
entirely different inference stacks (a local 4-bit quantized model via
Ollama versus a commercial API). This raises the possibility that the
direction of zero-shot prediction imbalance is linked to dataset
characteristics rather than being an idiosyncrasy of any one model.
Confirming that hypothesis would require testing additional model families
under the primary 3-seed protocol.

Cost differs substantially by deployment mode. At the Anthropic rates
used in our August 2026 evaluation, Claude Haiku 4.5 incurs a reported
API expenditure of \$0.94 per 1{,}000 incidents under zero-shot
prompting and \$2.15 per 1{,}000 incidents under RAG. The additional
retrieved context therefore increases the reported API expenditure by
2.29$\times$, corresponding to an additional \$1.21 per 1{,}000
incidents. In contrast, the locally served Qwen2.5-14B,
Mistral-Small-22B, and Llama-3.1-70B configurations incur no per-call
API fee in our comparison. This should not be interpreted as zero total
cost: GPU acquisition, power, cooling, maintenance, utilization, and
amortization are outside the scope of the reported API-cost analysis.
Table~\ref{tab:cost-summary} summarizes the cost basis of every model
and support track included in the revised benchmark. For a deployment
processing incidents continuously, the API expenditure can accumulate
with workload volume, while our batching ablation
(Table~\ref{tab:batch-sweep}) shows that a single RTX PRO 6000 can sustain
substantially higher throughput under batching on the tested workload.
We treat this as a single-provider, single-model preliminary comparison
rather than a definitive cost analysis; a broader comparison across
multiple cloud providers and model tiers is planned as future work
(Section~\ref{sec:limitations}).

\begin{table}[t]
\centering

\caption{Cost-aware summary of the model and support tracks evaluated in TriCalRAG.}
\label{tab:cost-summary}

\small
\setlength{\tabcolsep}{3.5pt}
\renewcommand{\arraystretch}{1.15}

\begin{tabularx}{\columnwidth}{
    >{\raggedright\arraybackslash}p{2.00cm}
    >{\raggedright\arraybackslash}p{2.55cm}
    >{\centering\arraybackslash}p{1.00cm}
    >{\centering\arraybackslash}p{1.00cm}
    >{\centering\arraybackslash}p{2.15cm}
    >{\centering\arraybackslash}p{0.55cm}
    >{\centering\arraybackslash}X
}

\toprule

\textbf{Track}
&
\textbf{Model}
&
\textbf{Deploy}
&
\textbf{Precision}
&
\textbf{Prompting}
&
\textbf{Seeds}
&
\textbf{Cost}
\\

\midrule

Primary
&
Qwen2.5-14B
&
Local
&
bf16
&
ZS + FS + RAG
&
3
&
No API fee
\\

Primary
&
Mistral-Small-22B
&
Local
&
bf16
&
ZS + FS + RAG
&
3
&
No API fee
\\

Supplement
&
Llama-3.1-70B
&
Local
&
4-bit
&
ZS + RAG
&
1
&
No API fee
\\

Supplement
&
Claude Haiku 4.5
&
Cloud
&
API
&
ZS + RAG
&
1
& 
\hspace{0.15cm}\$0.94 / 1K
ZS;\newline \$2.15 / 1K RAG
\\

Baseline
&
DeepLog
&
Local
&
LSTM
&
N/A
&
1
&
Not measured
\\

RAG support
&
MiniLM-L6-v2
+ FAISS
&
Local
&
FP
&
Top-3
&
N/A
&
Not measured
\\

\bottomrule

\end{tabularx}

\vspace{1mm}

\noindent
\begin{minipage}{\columnwidth}
\footnotesize
FP: Full Precision, bf16: Brain Floating Point 16-bit, 4-bit: 4-bit quantized precision, ZS: Zero-Shot, FS: Few-Shot, RAG: Retrieval-Augmented Generation. Note: All local models were served without a per-call API charge. Local hardware, power, cooling, maintenance, utilization, and amortization were not separately measured. Claude costs represent the reported API expenditure per 1K incidents.
\end{minipage}

\end{table}

The cost comparison in Table~\ref{tab:cost-summary} therefore separates
measured API expenditure from unmeasured local infrastructure cost.
Only the Claude track supports a direct monetary comparison in the
current study. The remaining tracks should be interpreted in terms of
their serving and computational requirements rather than as having zero
economic cost. A complete total-cost-of-ownership analysis would require
measuring hardware amortization, energy consumption, utilization,
maintenance, and operational overhead over a defined deployment period.

\section{Limitations}
\label{sec:limitations}

This section outlines the key limitations of the benchmark and the scope
of its findings.

\textbf{Windowing granularity.}
Our fixed five-line contiguous windowing scheme, applied uniformly across
all four datasets, is well matched to BGL and Thunderbird, where anomaly
labels are associated with localized log events, but is less well matched
to HDFS, whose anomaly labels are defined at the block level and whose
related log lines may occur non-contiguously. This mismatch likely
contributes to the lower HDFS performance observed across models and
prompting strategies. Block-aware incident construction should therefore
be considered in future evaluations.

\textbf{JSON extraction.}
An initial version of our output parser stripped only markdown code fences
at the beginning of a response. Some Mistral-Small responses contained
additional text before the JSON object, causing valid outputs to be
misclassified as parse failures. We corrected this by extracting the text
enclosed by the first opening curly brace (\texttt{\{}) and the last
closing curly brace (\texttt{\}}) before parsing it as JSON, reducing the
overall parse-failure rate to 0.2\%. We report this issue because it
materially affected the early results and illustrates the importance of
validating structured-output parsers across different model families.

\textbf{Calibration-baseline scoring.}
The Contextual Calibration and Batch Calibration experiments use
next-token log-probabilities, whereas the primary benchmark evaluates
decisions extracted from structured JSON outputs. These are distinct
decision paths and their results are therefore not numerically
interchangeable. The calibration comparison should be interpreted as an
internally consistent mechanism check on label-bias correction rather
than as a restatement of the main F1 results.

\textbf{DeepLog evaluation scale.}
The DeepLog baseline is trained and evaluated on a smaller held-out sample
than the LLM configurations. DeepLog uses 240 training sequences and
120 held-out evaluation incidents, whereas each LLM evaluation run
contains 600 incidents. Because DeepLog requires an explicit training
split, its evaluation protocol differs from that of the LLMs. The
reported DeepLog result should therefore be interpreted as a classical
reference baseline rather than as a directly paired model comparison.
A larger-scale classical evaluation may produce different results.

\textbf{Two-model primary scope.}
The repeated three-seed primary protocol evaluates two open-weight models:
Qwen2.5-14B and Mistral-Small. We additionally report single-run
supplementary checks with a 70B-class quantized model and a cloud-served
model. These experiments provide supplementary evidence that similar
prediction-balance behavior can occur in additional model and deployment
settings, but they use a different repetition protocol and should not be
interpreted with the same statistical confidence as the primary
three-seed results. Broader repeated evaluation across additional model
sizes and architecture families, including mixture-of-experts models,
remains future work.

\textbf{Explanation quality.}
The primary quantitative evaluation scores only the binary
\texttt{is\_anomaly} decision. Although the LLM-based systems also
generate \texttt{root\_cause} and \texttt{remediation} fields, we do not
quantitatively evaluate their semantic correctness, usefulness, or
faithfulness in the current study. Such evaluation would require an
additional human-annotation or carefully validated model-based judging
protocol. Consequently, F1 should not be interpreted as evidence of the
quality of the generated root-cause or remediation text.

\section{Conclusion and Future Work}
\label{sec:conclusion}

We presented TriCalRAG, a benchmark for on-premise LLM-based root
cause analysis evaluated entirely on a single workstation GPU. Across four
real log datasets, two open-weight models, and three prompting strategies,
we find that retrieval-augmented prompting consistently improves detection
performance and substantially reduces extreme prediction imbalance.
Seven of eight RAG configurations remain within the predefined
predicted-positive-rate range, compared with only two of eight zero-shot
configurations. This matters practically because a model that predicts
``anomaly'' on nearly every incident may still achieve a moderate F1 score
while providing limited discriminative value.

We also find that different models exhibit different operational
trade-offs: Mistral-Small achieves higher macro-averaged F1
(0.644 vs.\ 0.560), while Qwen2.5-14B exhibits fewer extreme
prediction-balance cases and approximately twice the throughput. The
held-out DeepLog baseline attains F1 = 0.698, numerically above the
macro-averaged F1 of both LLMs. However, because DeepLog is evaluated on
a smaller held-out set under a different training and evaluation
protocol, these values should be interpreted as complementary rather
than as a direct model-to-model ranking. Unlike DeepLog, the LLM-based
systems additionally generate natural-language root-cause and remediation
fields, whose semantic quality is not quantitatively evaluated in the
current study.

To examine whether the observed prediction-balance behavior is explained
by a known in-context label-bias mechanism, we compare RAG against two
published calibration-correction methods: Contextual
Calibration~\cite{zhao2021calibrate} and Batch
Calibration~\cite{zhou2024batch}. Neither consistently reproduces the
reduction in prediction imbalance observed with RAG. Supplementary
single-run checks with Llama-3.1-70B and Claude Haiku 4.5 show similar
dataset-dependent prediction patterns, with over-prediction on BGL and
Thunderbird and under-prediction on HDFS and OpenStack. These results
suggest that the observed behavior may depend partly on dataset
characteristics, although confirming this requires broader repeated
evaluation across additional model families.

Our ablations support the feasibility of the single-GPU setting:
batching increases throughput by approximately 41$\times$ (48 to 1991
tokens/s from batch size 1 to 128), while AWQ 4-bit quantization reduces
latency by approximately 20\% without measurable degradation on the
tested workload. These results indicate that single-GPU deployment is
practical for the evaluated workload, while prompt design and prediction
behavior remain important operational considerations.

Several directions remain. First, broader three-seed evaluation across
additional open-weight architectures and cloud providers would strengthen
the generality of the current findings. Second, block-aware windowing for
HDFS should address the evaluation-unit mismatch identified in this
study. Third, evaluating the semantic quality of generated root-cause
and remediation text would assess the capability that most clearly
distinguishes LLM-based RCA from classical anomaly detection, although
doing so rigorously requires an appropriate human- or model-based
evaluation protocol.

Finally, we explore extending TriCalRAG beyond application logs to
hardware-level signals, including GPU telemetry, Baseboard Management
Controller (BMC) sensors, boot-trust events, and bare-metal provisioning
data. These extensions use synthetic data and are therefore presented
only as directions for future work, not as validated experimental results.

\section*{Reproducibility}
Code, dataset splits, and evaluation harness are publicly available at:
\url{https://github.com/SPriTLab-CSE-IITJ/TriCalRAG}

\begin{acks}
The authors used ChatGPT-5.6 only for grammatical revision of the text in the paper to correct any typos, grammatical errors, and awkward phrasing. This work was supported by the Indian Institute of Technology Jodhpur, India under the Research Initiation Grant (RIG) Program (Grant No. I/I/RIG/SKM/20250216).
\end{acks}

\bibliographystyle{ACM-Reference-Format}
\bibliography{reference}

@inproceedings{zhao2021calibrate,
  title     = {Calibrate Before Use: Improving Few-Shot Performance of Language Models},
  author    = {Zhao, Zihao and Wallace, Eric and Feng, Shi and Klein, Dan and Singh, Sameer},
  booktitle = {Proceedings of the 38th International Conference on Machine Learning},
  pages     = {12603--12613},
  year      = {2021},
  editor    = {Meila, Marina and Zhang, Tong},
  volume    = {139},
  series    = {Proceedings of Machine Learning Research},
  publisher = {PMLR},
  url       = {https://proceedings.mlr.press/v139/zhao21c.html}
}

@inproceedings{zhou2024batch,
  title={Batch calibration: Rethinking calibration for in-context learning and prompt engineering},
  author={Zhou, Han and Wan, Xingchen and Proleev, Lev and Mincu, Diana and Chen, Jilin and Heller, Katherine and Roy, Subhrajit},
  booktitle={International Conference on Learning Representations},
  volume={2024},
  pages={49--70},
  year={2024}
}

@inproceedings{vllm2023,
  title={Efficient Memory Management for Large Language Model Serving with PagedAttention},
  author={Kwon, Woosuk and Li, Zhuohan and Zhuang, Siyuan and Sheng, Ying and Zheng, Lianmin and Yu, Cody Hao and Gonzalez, Joseph E and Zhang, Hao and Stoica, Ion},
  booktitle={Proceedings of the 29th Symposium on Operating Systems Principles},
  year={2023}
}

@inproceedings{loghub,
  title={LogHub: A Large Collection of System Log Datasets for AI-driven Log Analytics},
  author={Zhu, Jieming and He, Shilin and Liu, Jinyang and He, Pinjia and Xie, Qi and Zheng, Zibin and Lyu, Michael R},
  booktitle={Proceedings of the 34th International Symposium on Software Reliability Engineering (ISSRE)},
  year={2023}
}

@inproceedings{deeplog2017,
  title={DeepLog: Anomaly Detection and Diagnosis from System Logs through Deep Learning},
  author={Du, Min and Li, Feifei and Zheng, Guineng and Srikumar, Vivek},
  booktitle={Proceedings of the 2017 ACM SIGSAC Conference on Computer and Communications Security},
  year={2017}
}

@inproceedings{loganomaly2019,
  title={LogAnomaly: Unsupervised Detection of Sequential and Quantitative Anomalies in Unstructured Logs},
  author={Meng, Weibin and Liu, Ying and Zhu, Yichen and Zhang, Shenglin and Pei, Dan and Liu, Yuchi and Chen, Yihao and Zhang, Ruizhi and Tao, Shimin and Sun, Pei and others},
  booktitle={IJCAI},
  year={2019}
}

@article{loglizer,
  title={Tools and Benchmarks for Automated Log Parsing},
  author={He, Pinjia and Zhu, Jieming and He, Shilin and Li, Jian and Lyu, Michael R},
  journal={International Conference on Software Engineering: Software Engineering in Practice},
  year={2019}
}

@article{rohitpatel2026trishieldrag,
  title={TriShieldRAG: A Three-Ring Defense-in-Depth Framework Against Knowledge Corruption in Retrieval-Augmented Generation},
  author={Mohanty, Susil Kumar and Patel, Rohit and Yuvaraj, Kosuru and Chaudhary, Jeenal and Singhania, Disha},
  journal={arXiv preprint arXiv:2607.23838},
  eprint={2607.23838},
  archivePrefix={arXiv},
  primaryClass={cs.CR},
  url={https://arxiv.org/abs/2607.23838},
  year={2026}
}

@inproceedings{lograg2024,
  title={Leveraging RAG-Enhanced Large Language Model for Semi-Supervised Log Anomaly Detection},
  author={Zhang, Wanhao and Zhang, Qianli and Yu, Enyu and Ren, Yuxiang and Meng, Yeqing and Qiu, Mingxi and Wang, Jilong},
  booktitle={IEEE International Symposium on Software Reliability Engineering (ISSRE)},
  pages={168--179},
  year={2024}
}

@inproceedings{loggpt2023,
  title={LogGPT: Exploring ChatGPT for Log-Based Anomaly Detection},
  author={Qi, Jiaxing and Huang, Shaohan and Luan, Zhongzhi and Yang, Shu and Fung, Carol and Yang, Hailong and Qian, Depei and Shang, Jing and Xiao, Zhiyin and Wu, Zhihui},
  booktitle={IEEE International Conference on High Performance Computing \& Communications (HPCC)},
  pages={273--280},
  year={2023}
}

@article{logllm2024,
  title={LogLLM: Log-based Anomaly Detection Using Large Language Models},
  author={Guan, Wei and Cao, Jian and Qian, Shiyou and Gao, Jianwei and Ouyang, Chun},
  journal={arXiv preprint arXiv:2411.08561},
  year={2024}
}

@inproceedings{gpt4rca2024,
  title={Automated Root Causing of Cloud Incidents Using In-Context Learning with GPT-4},
  author={Zhang, Xuchao and Ghosh, Supriyo and Bansal, Chetan and Wang, Rujia and Ma, Minghua and Kang, Yu and Rajmohan, Saravan},
  booktitle={Companion Proceedings of the 32nd ACM International Conference on the Foundations of Software Engineering},
  pages={266--277},
  year={2024}
}

@article{llmeventlogsurvey2025,
  title={LLM-based Event Log Analysis Techniques: A Survey},
  author={Akhtar, Siraaj and Khan, Saad and Parkinson, Simon},
  journal={arXiv preprint arXiv:2502.00677},
  year={2025}
}

@article{hu2022lora,
  title={LoRA: Low-Rank Adaptation of Large Language Models},
  author={Hu, Edward J and Shen, Yelong and Wallis, Phillip and Allen-Zhu, Zeyuan and Li, Yuanzhi and Wang, Shean and Wang, Lu and Chen, Weizhu},
  journal={International Conference on Learning Representations},
  year={2022}
}

@article{dao2022flashattention,
  title={FlashAttention: Fast and Memory-Efficient Exact Attention with IO-Awareness},
  author={Dao, Tri and Fu, Daniel Y and Ermon, Stefano and Rudra, Atri and R{\'e}, Christopher},
  journal={Advances in Neural Information Processing Systems},
  volume={35},
  pages={16344--16359},
  year={2022}
}

@article{lin2024awq,
  title={AWQ: Activation-aware Weight Quantization for LLM Compression and Acceleration},
  author={Lin, Ji and Tang, Jiaming and Tang, Haotian and Yang, Shang and Chen, Wei-Ming and Wang, Wei-Chen and Xiao, Guangxuan and Dang, Xingyu and Gan, Chuang and Han, Song},
  journal={Proceedings of Machine Learning and Systems},
  volume={6},
  year={2024}
}

@article{kaplan2020scaling,
  title={Scaling Laws for Neural Language Models},
  author={Kaplan, Jared and McCandlish, Sam and Henighan, Tom and Brown, Tom B and Chess, Benjamin and Child, Rewon and Gray, Scott and Radford, Alec and Wu, Jeffrey and Amodei, Dario},
  journal={arXiv preprint arXiv:2001.08361},
  year={2020}
}

@techreport{qwen2024technical,
  title={Qwen2.5 Technical Report},
  author={{Qwen Team}},
  institution={Alibaba Group},
  journal={arXiv preprint arXiv:2412.15115},
  year={2024}
}

@article{hendrycks2021mmlu,
  title={Measuring Massive Multitask Language Understanding},
  author={Hendrycks, Dan and Burns, Collin and Basart, Steven and Zou, Andy and Mazeika, Mantas and Song, Dawn and Steinhardt, Jacob},
  journal={International Conference on Learning Representations},
  year={2021}
}

@article{liang2022helm,
  title={Holistic Evaluation of Language Models},
  author={Liang, Percy and Bommasani, Rishi and Lee, Tony and Tsipras, Dimitris and Soylu, Dilara and Yasunaga, Michihiro and Zhang, Yian and Narayanan, Deepak and Wu, Yuhuai and Kumar, Ananya and others},
  journal={Transactions on Machine Learning Research},
  year={2023}
}

@article{jimenez2023swebench,
  title={SWE-bench: Can Language Models Resolve Real-World GitHub Issues?},
  author={Jimenez, Carlos E and Yang, John and Wettig, Alexander and Yao, Shunyu and Pei, Kexin and Press, Ofir and Narasimhan, Karthik},
  journal={International Conference on Learning Representations},
  year={2024}
}

@article{efron1979bootstrap,
  title={Bootstrap Methods: Another Look at the Jackknife},
  author={Efron, Bradley},
  journal={The Annals of Statistics},
  volume={7},
  number={1},
  pages={1--26},
  year={1979}
}

@article{chandola2009anomaly,
  title={Anomaly Detection: A Survey},
  author={Chandola, Varun and Banerjee, Arindam and Kumar, Vipin},
  journal={ACM Computing Surveys},
  volume={41},
  number={3},
  pages={1--58},
  year={2009}
}

@article{soldani2022rca,
  title={Anomaly Detection and Failure Root Cause Analysis in (Micro) Service-Based Cloud Applications: A Survey},
  author={Soldani, Jacopo and Brogi, Antonio},
  journal={ACM Computing Surveys},
  volume={55},
  number={3},
  pages={1--39},
  year={2022}
}

@inproceedings{dang2019aiops,
  title={AIOps: Real-World Challenges},
  author={Dang, Yingnong and Lin, Qingwei and Huang, Peng},
  booktitle={IEEE/ACM 41st International Conference on Software Engineering: Software Engineering in Practice (ICSE-SEIP)},
  pages={4--5},
  year={2019}
}

@inproceedings{ahmed2023recommending,
  title={Recommending Root-Cause and Mitigation Steps for Cloud Incidents Using Large Language Models},
  author={Ahmed, Toufique and Ghosh, Supriyo and Bansal, Chetan and Zimmermann, Thomas and Zhang, Xuchao and Rajmohan, Saravan},
  booktitle={Proceedings of the 45th International Conference on Software Engineering (ICSE)},
  pages={1737--1749},
  year={2023}
}

@article{li2025anomalygen,
  title={Anomaly-Gen: An Automated Semantic Log Sequence Generation Framework with LLM for Anomaly Detection},
  author={Li, Xiao and Huo, Yintong and Mao, Chenxi and Shan, Shiwen and Su, Yuxin and Li, Dan and Zheng, Zibin},
  journal={arXiv preprint arXiv:2504.12250},
  year={2025}
}

@inproceedings{he2024llmelog,
  title={LLMeLog: An Approach for Anomaly Detection Based on LLM-Enriched Log Events},
  author={He, Minghua and Jia, Tong and Duan, Chiming and Cai, Huaqian and Li, Ying and Huang, Gang},
  booktitle={IEEE International Symposium on Software Reliability Engineering (ISSRE)},
  pages={132--143},
  year={2024}
}

@article{liu2024loglm,
  title={LogLM: From Task-based to Instruction-based Automated Log Analysis},
  author={Liu, Yilun and Ji, Yuhe and Tao, Shimin and He, Minggui and Meng, Weibin and Zhang, Shenglin and Sun, Yongqian and Xie, Yuming and Chen, Bo and Yang, Hao},
  journal={arXiv preprint arXiv:2410.09352},
  year={2024}
}

@inproceedings{duan2025eagerlog,
  title={EagerLog: Active Learning Enhanced Retrieval Augmented Generation for Log-Based Anomaly Detection},
  author={Duan, Chiming and Jia, Tong and Yang, Ying and Liu, Guoyao and Liu, Jia and Zhang, Huan and Zhou, Qi and Li, Ying and Huang, Gang},
  booktitle={IEEE International Conference on Acoustics, Speech and Signal Processing (ICASSP)},
  pages={1--5},
  year={2025}
}

@inproceedings{xiao2025clslog,
  title={ClsLog: Collaborating Large and Small Models for Log-Based Anomaly Detection},
  author={Xiao, Pengxiang and Jia, Tong and Duan, Chiming and He, Minghua and Hong, Wenjun and Yang, Xinyu and Wu, Ying and Li, Ying and Huang, Gang},
  booktitle={Companion Proceedings of the 33rd ACM International Conference on the Foundations of Software Engineering (FSE Companion)},
  pages={686--690},
  year={2025}
}

@article{jiang2024lilac,
  title={Lilac: Log Parsing Using LLMs with Adaptive Parsing Cache},
  author={Jiang, Zhihan and Liu, Jinyang and Chen, Zhuangbin and Li, Yichen and Huang, Junjie and Huo, Yintong and He, Pinjia and Gu, Jiazhen and Lyu, Michael R},
  journal={Proceedings of the ACM on Software Engineering (PACMSE)},
  volume={1},
  year={2024}
}

@inproceedings{song2025confront,
  title={Confront Insider Threat: Precise Anomaly Detection in Behavior Logs Based on LLM Fine-Tuning},
  author={Song, Shuang and Zhang, Yifei and Gao, Neng},
  booktitle={Proceedings of the 31st International Conference on Computational Linguistics (COLING)},
  pages={8589--8601},
  year={2025}
}

@inproceedings{xu2024unilog,
  title={UniLog: Automatic Logging via LLM and In-Context Learning},
  author={Xu, Junjielong and Cui, Ziang and Zhao, Yuan and Zhang, Xu and He, Shilin and He, Pinjia and Li, Liqun and Kang, Yu and Lin, Qingwei and Dang, Yingnong and others},
  booktitle={Proceedings of the 46th IEEE/ACM International Conference on Software Engineering (ICSE)},
  pages={1--12},
  year={2024}
}

@inproceedings{zhong2024logparserllm,
  title={LogParser-LLM: Advancing Efficient Log Parsing with Large Language Models},
  author={Zhong, Aoxiao and Mo, Dengyao and Liu, Guoxin and Xie, Jing and Chen, Yudong and Zhang, Qingnan and Xu, Xinyi and Zhang, Bo and Wang, Wei and Xie, Yi and others},
  booktitle={Proceedings of the 30th ACM SIGKDD Conference on Knowledge Discovery and Data Mining},
  pages={4559--4570},
  year={2024}
}

@inproceedings{lewis2020rag,
  title={Retrieval-Augmented Generation for Knowledge-Intensive NLP Tasks},
  author={Lewis, Patrick and Perez, Ethan and Piktus, Aleksandra and Petroni, Fabio and Karpukhin, Vladimir and Goyal, Naman and K{\"u}ttler, Heinrich and Lewis, Mike and Yih, Wen-tau and Rockt{\"a}schel, Tim and Riedel, Sebastian and Kiela, Douwe},
  booktitle={Advances in Neural Information Processing Systems},
  volume={33},
  pages={9459--9474},
  year={2020}
}

@article{johnson2019faiss,
  title={Billion-Scale Similarity Search with {GPUs}},
  author={Johnson, Jeff and Douze, Matthijs and J{\'e}gou, Herv{\'e}},
  journal={IEEE Transactions on Big Data},
  volume={7},
  number={3},
  pages={535--547},
  year={2019}
}

@inproceedings{reimers2019sbert,
  title={Sentence-{BERT}: Sentence Embeddings using {S}iamese {BERT}-Networks},
  author={Reimers, Nils and Gurevych, Iryna},
  booktitle={Proceedings of the 2019 Conference on Empirical Methods in Natural Language Processing (EMNLP-IJCNLP)},
  pages={3982--3992},
  year={2019}
}

@misc{mistralsmall2409,
  title        = {Mistral-Small-Instruct-2409},
  author       = {{Mistral AI Team}},
  year         = {2024},
  howpublished = {Hugging Face model card},
  url          = {https://huggingface.co/mistralai/Mistral-Small-Instruct-2409},
  note         = {22B instruction-tuned model}
}

\end{document}